\RequirePackage{fix-cm}
\documentclass[smallextended]{svjour3}       % onecolumn (second format)
\smartqed  % flush right qed marks, e.g. at end of proof
\usepackage{graphicx}
\usepackage{amsmath}
\usepackage{booktabs} % For formal tables

\usepackage{adjustbox}
\usepackage{xspace}
\usepackage{multirow}
\usepackage{hhline}
\usepackage{hyperref}
\usepackage{balance}

\usepackage{xcolor}

\PackageWarning{TODO:}{X}
\PackageWarning{TODO:}{X\%}

\newcommand{\etal}{\hbox{\emph{et al.}}\xspace}
\newcommand{\eg}{\hbox{\emph{e.g.,}}\xspace}
\newcommand{\ie}{\hbox{\emph{i.e.}}\xspace}

\begin{document}

%\title{The Impact of Code Review Measures \\ on Predicting Post-Release Defects}
\title{Do Code Review Measures Explain the Incidence of Post-Release Defects?}
%\titlerunning{The Impact of Code Review Measures on Post-Release Defects}        % if too long for running head
\subtitle{Case Study Replications and Bayesian Networks}

\author{Andrey Krutauz         \and
        Tapajit Dey          \\  \and 
        Peter C. Rigby \and
        Audris Mockus        
}


\institute{
			Andrey Krutauz \\
			Concordia University  \at
              Montreal, QC, Canada \\
              \email{andrey.krutauz@ensce.concordia.ca} 
           \and
           Tapajit Dey  \\ 
           University of Tennessee \at
             Knoxville, Tennessee, USA \\
             \email{tdey2@vols.utk.edu}
             \and
             Peter C. Rigby \\
             Concordia University  \at
              Montreal, QC, Canada \\
              \email{peter.rigby@concordia.ca} 
           \and
           Audris Mockus        \\
           University of Tennessee \at
             Knoxville, Tennessee, USA \\
             \email{audris@utk.edu}
}

\date{Received: date / Accepted: date}

\maketitle

\begin{abstract}
Aim: In contrast to studies of defects found during code review, we
aim to clarify whether code reviews measures can explain the prevalence of 
post-release defects.

Method: We replicate McIntosh \etal's~\cite{mcintosh2015emse} study
that uses additive regression to model the relationship between
defects and code reviews. To increase 
external validity, we apply the same methodology on a new software
project. We discuss our findings with the first author of the
original study, McIntosh. We then investigate how to reduce the
impact of 
correlated predictors in the variable selection process and how to
increase understanding of the inter-relationships among the
predictors by employing Bayesian Network (BN) models.

Context: As in the original study, we use the same measures authors
obtained for Qt project in the original study. We mine data from
version control and issue tracker of Google Chrome and operationalize
measures that are close analogs to the large collection of code,
process, and code review measures used in the replicated
the study.

Results: Both the data from the original study and the Chrome data showed high instability of the influence of code review measures on defects with
the results being highly sensitive to variable selection
procedure. Models without code review predictors had as good or
better fit than those with review predictors. Replication, however,
confirms with the bulk of prior work showing that prior defects,
module size, and authorship have the strongest relationship to
post-release defects. The application of BN models helped explain
the observed instability by demonstrating that the review-related
predictors do {\it not} affect post-release defects directly and showed indirect
effects. For example, changes that have \emph{no review discussion}
tend to be associated with files that have had many \emph{prior
  defects} which in turn increase the number of post-release
defects.
We hope that similar analyses of other software engineering techniques may also yield a more nuanced view of their impact. Our replication package including our data and scripts is publicly available~\cite{ReplicationPackage}.
%\and  visual models}
\end{abstract}

\section{Introduction}
\label{intro}
For decades code review has been seen as a cornerstone of quality
assurance for software projects. The process evolved from a formal
process with checklists and face to face
meetings~\cite{fagan2002history} to a lightweight and semi-formal
review done via e-mails or specially designed collaboration
tools~\cite{rigby2011understanding}. The lightweight code review
approach was originally used in open source software projects (OSS),
because of their highly distributed
nature~\cite{mockus2000case,Rigby2008ICSE} and has also become a
common practice among commercial projects as well
\cite{rigby2013convergent,bacchelli2013expectations}. Recent
studies suggest that the focus of review has shifted from early
defect discovery to problem discussion and knowledge
sharing~\cite{bacchelli2013expectations,rigby2013convergent,Bosu2015MSR,Rahman2017MSR,Kononenko2016ICSE}. It
is perceived as a major quality control mechanism to
prevent defects in production code~\cite{bacchelli2013expectations,Meneely2015MSR,Meneely2017EMSE}.

An obvious and important scientific question is whether or not
code reviews actually improve software quality, and whether our measurements of code review have explanatory power. To clarify such
theoretical question, science resorts to replication to make it
self-correcting system~\cite{shull2008role,carver1978case}. Replication helps establish 
if the phenomenon is dependable or
idiosyncratic~\cite{runeson2009guidelines,shull2002replicating,axelrod1997advancing,Carver2010Replication}. Our
first aim is, therefore to conduct a similar-internal
replication (a replication where only the experimenters 
varied~\cite{gomez2014understanding,almqvist2006replication}) of a highly reputable
recent result investigating the effects code reviews have on software
quality.  We chose a commonly used quality measure: post-release
defects. Such defects affect end-users (and vendor reputation) and
are very costly to repair~\cite{nist02}, and, therefore are a primary
concern to software industry~\cite{huang2006much}.

\textbf{RQ 1. Replication: do previously reported associations
  between code review measures and post-release defects hold
  in a similar-internal replication study?}

To investigate a hypothesis-driven scientific question researchers
often use linear regression models~\footnote{Machine learning
  methods  focused on maximising prediction performance are widely used for defect prediction, but such methods
  are typically not transparent enough to test scientific
  hypothesis~\cite{lin2013research}} to examine the relation between code review (and
other metrics) and software
quality~\cite{Porter1998TOSEM,kononenko2015investigating,morales2015code}. A
recent award-winning work by McIntosh \emph{et al.}
~\cite{mcintosh2014impact,mcintosh2015emse} employed additive models to fit non-linear
curves that are more suited for non-monotone or non-linear
relationships than linear regression. We perform an exact replication
of that experiment to determine if we can obtain the same
conclusions using the same methods and data. Specifically, we
construct OLS models with restricted cubical splines to model these
relationships and discuss our findings with the first author, McIntosh, of this
study to ensure that he agreed with our conclusions.

\textbf{RQ 2. Differentiated-external
replication: do previously reported associations
  between code review measures and post-release defects hold
  for another large software project?}  
  
Our second goal is to increase
external validity~\cite{gomez2014understanding} of the results to
avoid conclusions that are unique to the specific dataset reported
in the paper. To accomplish this, we apply exactly the same set of
methods on a different software project: Chrome. This is sometimes
referred as differentiated-external
replication~\cite{almqvist2006replication}.  We chose the project
due to its size and richness and quality of the associated data that
allowed us to obtain measures highly similar to ones obtained in the
Qt project of the replicated study. More specifically, we model
software defects that are reported in a bug tracker. As control
variables, we use many of the previously studied measures that have
been shown to impact defects, including size, complexity, churn,
authors, and file
ownership~\cite{bird2011don,hassan2009predicting}. The focus of this
study is on investigating code review measures many of which have been examined in
past studies, including the number of reviewers, discussion length,
and rushed review in a different setting~\cite{mcintosh2015emse,mcintosh2014impact,kononenko2015investigating,Rigby2008ICSE,Rigby2014TOSEM}.

\textbf{RQ 3. Structure of the relationships: Are code review
  measures directly associated with post-release defects or are they
  affected by other measures of the development process that
  are, in turn, directly associated with post-release defects?}

The findings from RQ1 and RQ2 point to the methodological limitations of
linear regression and additive models when applied to datasets that
have high correlations among the predictors as is typical software
engineering data in general and in code review data in
particular~\cite{Rigby2014TOSEM,mcintosh2015emse}. The linear (or
additive) models can not reliably determine which of the highly
correlated predictors are affecting the response. Principal
Component Analysis (PCA) is typically applied in such cases but
the results are hard to interpret because a linear
combination of unrelated measures, e.g., combining lines of code,
number of reviewers, and other unrelated concepts into a single
predictor. This defeats the original purpose of testing the
scientific hypothesis as discussed in Chapters 6.3, 6.7, and 10.2
of~\cite{james2013introduction}. Since automatic variable selection
techniques are highly unstable (see, e.g,~\cite{vsel04}), best
practices in empirical studies that employ regression models, recommend the manual removal of highly correlated variables,
or variables that do not contribute to the explanatory power of the
model. Such selection of variables relies on a subjective judgement
of the researcher. Another shortcoming of such models is their
inability to model the relations among predictors, which may reveal
salient aspects of the development process by providing a rich
picture of how the predictors may influence each other and the
response. 

A Bayesian Network (BN) is a Probabilistic Graphical Model
(PGM). PGM describes probabilistic relationships among variables
that describe a problem domain~\cite{heckerman1998tutorial}. This
model has several advantages over linear or additive regression
models. In particular, it allows for a natural representation of
conditional dependence and independence using graph notation where
variables are nodes and dependencies are edges. The removal of the
notion of predictor and response variables disposes of the
oversimplifying assumption that a single response variable is
explained by a long list of predictors. Instead the edges in the
Bayesian Network provide a meaningful structure based on collected
data. Each variable in a graph can be interpreted as a predictor or
a response variable based on the topology of the graph. The
researcher can then inject information to understand the impact of
an edge of interest~\cite{friedman1999data}.

Our main findings from RQ 1, the reproduction of
the study by McIntosh \emph{et al.} ~\cite{mcintosh2015emse},
have demonstrated high sensitivity of the regression modeling results to the subjective
steps in the analyses when data contains highly correlated
predictors. In particular, we found that even in exact reproduction
we were unable to confirm the predictive power of code review measures on
post-release defects. We discussed the finding with McIntosh and,
according to his opinion, code review measures are not likely to explain more of the
variance than traditional measures. Moreover, the results are
inconsistent across 
software releases and heavily depend on the variables selected. We
did, however, find several metrics not related to code reviews,
such as churn or prior defects, that were reproduced reliably
despite the subjectivity of the variable selection process.

The investigation in RQ 2, increased the external
validity of the findings by confirming that a relatively small set
of measures, churn and prior defects, are related to post-release defects on a large and
unrelated software project. As on the Qt dataset, the impact of
review measures was inconsistent across software releases and
heavily depend on the variables selected.

In RQ 3, to reduce the subjectivity of variable selection process and to
untangle the complex web of dependencies among the predictors, we
applied Bayesian Networks (BN) on both datasets. The approach
revealed that there is no direct relation between review
measures and defects. The graph shows, for example, that modules
with more self-approved changes also have more changes with no
discussion, more reviewers, and also more prior defects. An increase in
review issues increases the share of the work done by the minor
authors, which, in turn, is associated with increased number of
defects. 

This paper is organized as follows. In Section \ref{secMethodology},
we discuss the case study design, the systems under study, and the data
extraction process. We also give a brief overview of the Chrome code
review process. In Section \ref{secReplication}, we replicate and reproduce 
McIntosh \emph{et al.}'s  \cite{mcintosh2015emse} study, describe the
model construction, results, and discussion.  In Section
\ref{secBN}, we describe BNs and discuss the findings from these
models. Threats to validity are discussed in Section
\ref{sec:limitations}. The final section concludes the paper and
suggests future work. Our replication package including our data and
scripts is publicly available~\cite{ReplicationPackage}.

\section{Case study design and data} 
\label{secMethodology}

In this section we discuss the case study design including the projects under study and reasons for their selection. We describe the data sources, steps in the data extraction, and analysis approach. We discuss the Bayesian Network modeling methodology in Section~\ref{secBN}. 

\subsection{Systems under study}

McIntosh \etal~\cite{mcintosh2015emse} mined code review data from Android,
LibreOffice, QT, ITK, and VTK. They did not conduct an analysis on Android and
LibreOffice because they found that many of the reviews were not linked to bug
reports which did not allow them to study the impact of review on bugs. In
total, they studied two QT releases and one release for VTK and ITK. For the
reproduction, McIntosh provided the Qt and ITK data that was used in their
work~\cite{mcintosh2015emse}. The ITK data had only 24 defective components and 344 commits with reviews. We feel that this dataset is too small to produce meaningful statistical models. Although we include the ITK results in our replication package~\cite{ReplicationPackage}, we only present the Qt results in this work. To improve external validity, we replicate the study on the Google Chrome
project, because like QT, it is large and primarily written in C++. A further reason for
studying Chrome is that it is an open source web browser, that is mostly
developed by paid Google developers and its development practices mirror those
used internally at Google. Chrome developers are required to perform code
review on each change and use Reitvield, the precursor to Gerrit, to improve
traceability of bugs, changes, and reviews.

For completeness, we briefly describe Chrome's code review process which resembles other modern review practices~\cite{rigby2012contemporary}.
A review begins when the change author submits a patch and invites
reviewers. A reviewer examines a change and either approves it by
replying with special keyword \emph{lgtm} (looks good to me) or
proposes improvements. The author addresses comments either by
fixing issues in code or by replying to the reviewer comments. Subsequent modifications to the original patch appear in the same review and are called \emph{patchsets}. The new patchset triggers a new cycle of review and revision. The process continues until all issues are fixed and the reviewers are satisfied with the patch. The code can then be merged to the trunk.

\subsection{Chrome data extraction}  

\begin{figure*}
%\centering
\includegraphics[width=.85\textwidth]{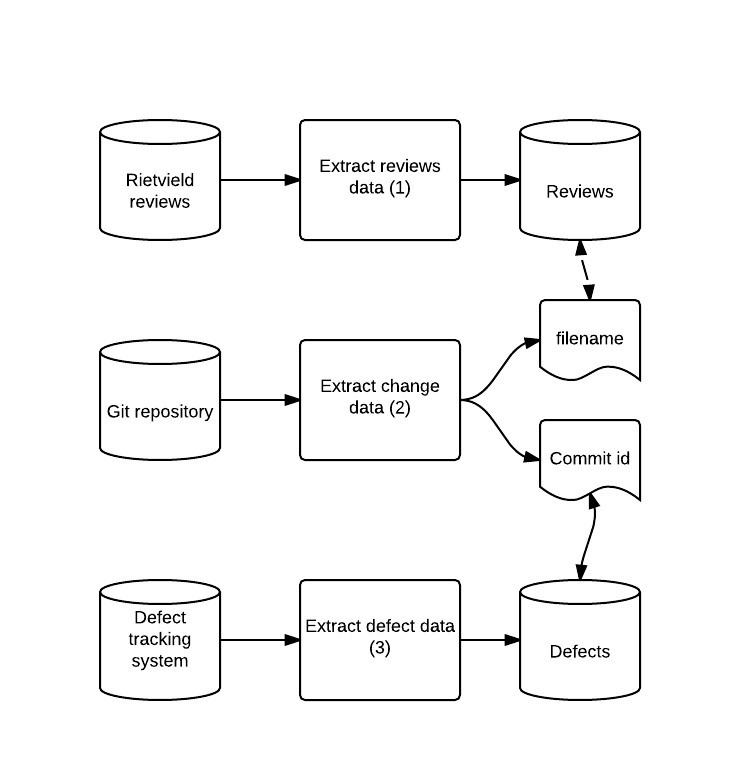}
\caption{Chrome data sources and extraction methodology. Reviews are extracted from the Rietrvield review system. Bugs are extracted from the defect tracking system. The bug and review ids are contained in the Git commit and linked to the modified files and directories.}
\label{fig_chrome_data_sources}
\end{figure*}

%\begin{table*}
\begin{table*}
\caption{Description of Measures: product, process, human factors, review participation, and reviewer expertise}
\label{tabMeasures}
\centering
% Some packages, such as MDW tools, offer better commands for making tables
% than the plain LaTeX2e tabular which is used here.
\begin{tabular}{l|p{.2\textwidth}|p{.55\textwidth}}

&Measure & Description\\
\hline
\hline
\multirow{2}{*}{Product} &
Size  & Number of lines of executable code in component \\
\hhline{~--}
\multirow{2}{*}{} & Complexity & The McCabe cyclomatic complexity. \\
\hline
\multirow{4}{*}{Process} & Prior defects & Number of defects fixed in component prior to the considered release period\\ 
\hhline{~--}
\multirow{4}{*}{} & Effective tests (Chrome only) & Total number of times a test found an issue during the review process \\
\hhline{~--}
\multirow{4}{*}{} & Churn & Sum of added or removed lines of code per component during considered period of time\\ 
\hhline{~--}
\multirow{4}{*}{} & Change entropy & Distribution of changes among files within a component\\ 
\hline
\multirow{4}{*}{Human Factors} & Minor authors & Number of unique contributors that contribute less than $5\%$ of code changes to a component\\
\hhline{~--}
\multirow{4}{*}{} & Major authors & Number of unique contributors that contribute at least $5\%$ of code changes to component \\
\hhline{~--}
\multirow{4}{*}{} & All authors & Number of unique contributors to component \\
\hhline{~--}
\multirow{4}{*}{} & Author ownership & Proportion of changes to component done by major authors\\
\hline
\multirow{9}{*}{Review Participation} & Rushed reviews & Number of reviews that were concluded faster than acceptable review rate (200 loc per hour)\\
\hhline{~--}
\multirow{9}{*}{} & Changes without discussion & Changes that were integrated without discussion comments\\
\hhline{~--}
\multirow{9}{*}{} & Self approved changes & Changes that were approved for integration only by the author\\
\hhline{~--}
\multirow{9}{*}{} & Typical discussion length & Discussion length typical for that specific component measured in number of discussion comments. Normalized by size of change (churn)\\
\hhline{~--}
\multirow{9}{*}{} & Typical review window & The amount of time between the patch upload and its approval for integration. Normalized by size of change (churn) \\
\hhline{~--}
\multirow{9}{*}{} & All reviews & Total number of times the component was reviewed \\
\hhline{~--}
\multirow{9}{*}{} & All reviewers & Total number of of reviewers that reviewed a component\\
\hhline{~--}
\multirow{9}{*}{} & Review issues & Total number of patch revisions created during review process \\
\hhline{~--}
\multirow{9}{*}{} & Effective reviews (Chrome only) & Number of revisions that led to a code change during a single review per component\\
\hline
\multirow{2}{*}{Review Expertise} & Lacking subject matter expertise & Number of changes that were not authored or approved by major author\\
\hhline{~--}
\multirow{2}{*}{} & Typical reviewer expertise & Total number of changes to the component authored or reviewed by this reviewer prior to this change  \\
\hline
\end{tabular}
\end{table*}

To understand the influence of code review measures on post release defects we need to create a link between the code review, the source files, and reported bugs. We collect data from three data sources: Reitvield, Git, and the Chrome bug tracker (figure \ref{fig_chrome_data_sources}). The data extraction is divided into the three steps described below.

{\it Extracting review data:} We use the Reitvield API to download code reviews in JSON format and extract the data into a database. For each code review patch revision we extract the unique identifier and the set of files modified by this revision. For every file and revision we also capture the number of added and removed lines to calculate the size of a change. We process the reviewers comments. We ignore comments that were added automatically by a bot or by the patch author. 

{\it Extracting Git repository information:} We extract commit information \ie the commit hash and list of files related to the change from the Git repository. We use the Understand static analysis toolkit\footnote{https://scitools.com/} to extract source code measures from the files.

{\it Extracting defect data:} We mine the defects from the Chrome issue tracker by scraping the pages. We extract the submission date, type of the issue, review ID, and commit ID for the fix.

{\it Post-release defects:} We consider a defect to be the post-release defect of the current release if it was submitted during the time period between the release dates of the current and the following releases.  We use Chrome release calendar website for release dates information.\footnote{https://www.chromium.org/developers/calendar} 
Following McIntosh~\cite{mcintosh2015emse}, we associate the
post-release defects with the pre-release reviews and other source
measures using first the file level and then sum the measures to the
component, \ie directory level. The directory was chosen as the unit of
analysis to reduce the fraction of zero observations because the
majority of the files in the system do not have any defects.

\subsection{Collected Measures}
The measures we collect to evaluate the impact of code review on post-release defects are well known and have been used in multiple past studies~\cite{mcintosh2015emse,mcintosh2014impact,rigby2013convergent,kononenko2015investigating} and are described in Table~\ref{tabMeasures}. We divide them into four categories: product, process, human factors, review participation, and reviewer expertise.
%~\footnote{We do not include review code coverage measures as these variables were determined to not consistently predict post release defects and were excluded from the study under reproduction} 
The code review measures are the number of reviewers, discussion length, rushed reviews, typical reviewer expertise, etc. The control variables in our model are also well known and widely used with defect prediction models~\cite{mockus2000case,hassan2009predicting,bird2011don,graves2000predicting}. The control variables include the size of the file, the number of prior defects, and the churn.

\section{Code Review Replication and Reproduction Study} \label{secReplication}

We replicate the study published by McIntosh \etal ~\cite{mcintosh2015emse}. We strictly follow the steps of the model construction described in the original paper~\cite{mcintosh2015emse}. We fit an Ordinary Least Squares (OLS) regression model. Since the dependent variable is the number of post-release defects and it is highly skewed, we log transform it. We also create regression models that take non-linear effects into account. We then compare the goodness of fit among models and discuss the contribution of each independent variable.

\begin{figure*}
    %\centering
   \includegraphics[angle=270,width=0.7\textwidth]{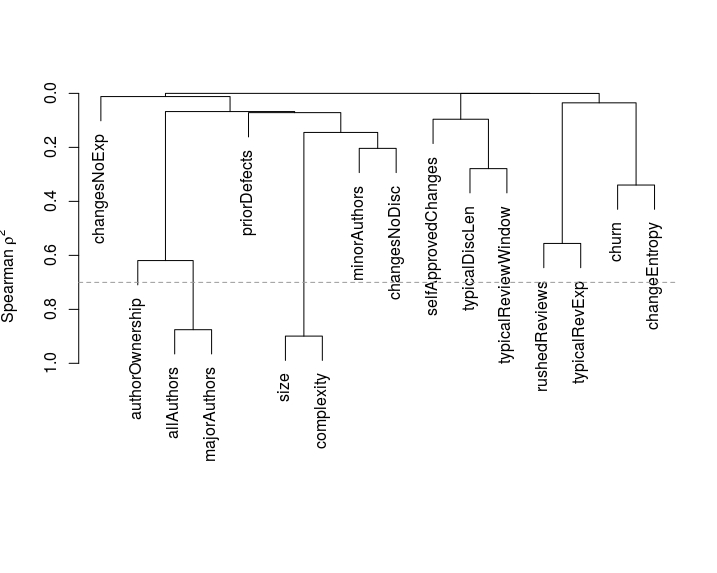}
   \caption{Hierarchical Correlation Analysis for Qt 5.0. For variables correlated at $|\rho|\geq0.7$ the simpler measure is kept. We also conduct a redundancy analysis. }
   \label{FigCorQt5}
   \includegraphics[angle=270,width=0.7\textwidth]{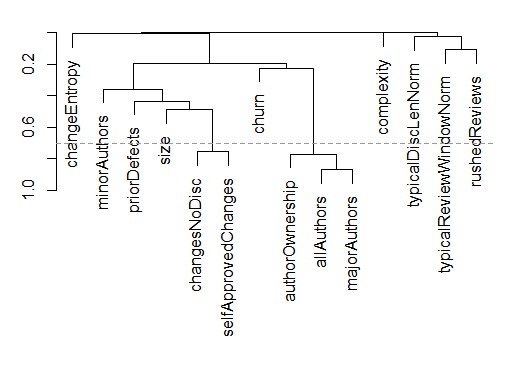}
   \caption{Hierarchical Correlation Analysis for Chrome 40. For variables correlated at $|\rho|\geq0.7$ the simpler measure is kept. We also conduct a redundancy analysis.}
   \label{FigCorChrome40}
\end{figure*}

{\it Correlated variables} can distort the contribution of a
variable to a model and must be removed. We use the hierarchical
clustering analysis (Figures~\ref{FigCorQt5} and
\ref{FigCorChrome40}) with the threshold of $|\rho|\geq0.7$
suggested by McIntosh~\cite{mcintosh2015emse} to identify the highly
correlated variables. Then we select which variables will be
discarded using drop-one analysis and the principle of parsimony. The
results of this step are summarized in the Tables~\ref{results_tbl}
and~\ref{results_tbl_chrome}. 

{\it Redundant variables} can be explained by the remaining variables, that is they do not contribute to the explanatory power of the model and should be removed. Such variables may be overlooked by the pairwise correlation analysis, therefore we use \emph{redun} function from \emph{Hmisc} R package~\footnote{\url{https://cran.r-project.org/web/packages/Hmisc/Hmisc.pdf}}.
For each independent variable a regression model is fitted using as
predictors the remaining variables. If the model has a $R^{2}$
greater than $0.9$, then the current variable is considered to be redundant because the
linear combination of the remaining variables can closely
approximate this variable. 

{\it Non-linear effects and degrees of freedom.} Traditional defect prediction models assume linear dependencies between the dependant and independent variables. McIntosh \etal \cite{mcintosh2015emse} showed that for some code review measures this relation has a non-linear shape. To identify variables that may be nonlinear, we calculate the Spearman multiple $\rho^{2}$ scores for each independent variable. Variables with higher scores are more likely to be non-linear. To fit a non-linear curve we use restricted cubic splines in the \emph{rms} R package~\cite{harrell2013rms}. Using this approach we assign knots, which are points where the slope changes, to potentially non-linear variables. The more knots that are added the greater the curve complexity. Every additional knot requires a degree of freedom. If we use all of the degrees of freedom, then there will be a knot for each data point and the fit will be perfect, but the model will be overfitted to the data. As a result, the degrees of freedom are budgeted to avoid over fitting while still allowing variables that have a high $\rho^{2}$ score to be modelled non-linearly. 

To {\it assess model fitness}, we report the adjusted $R^{2}$ to compensate for the large number of variables~\cite{mcintosh2015emse}. To assess the individual contributions of each variable, we report its statistical significance and the Wald $\chi^{2}$ maximum likelihood test value. The larger the value the greater the impact the variable has on the model. The results are summarized in Tables~\ref{results_tbl} and \ref{results_tbl_chrome}. The tables also contain the results from McIntosh \etal~\cite{mcintosh2015emse} and conform to the same structure.

\subsection{Variable selection and model construction}

In our summary table we have approximately 1.3k reviews for Qt 5.0  and 1.4K for Chrome 40. We start with 16 measures. This gives us 81 and 87 degrees of freedom for Qt and Chrome respectively. 
Following previous works, we discard measures with correlation at or above 0.7. We use clustering analysis to identify these measures, see Figures \ref{FigCorQt5} and \ref{FigCorChrome40}. We also perform \emph{drop one} analysis to determine which measures should be discarded from each cluster. \emph{Major authors}, \emph{author ownership} and \emph{complexity} were removed from Qt dataset. \emph{Major authors}, \emph{author ownership} and \emph{changes without discussion} were removed from Chrome. Using a redundancy test \emph{minor authors} was removed from both datasets. We perform a non-linearity analysis. Variables that exhibit a higher degree of non-linearity require additional degrees of freedom to model their curved line. 
The results of the variable selection process and the number of allocated degrees for each variable can be found in in Tables \ref{results_tbl} and \ref{results_tbl_chrome}.

To represent our models, we use the R language notation~\cite{pinheiro2011r}.  For example, the formula $y \sim a + b$ means that the response $y$ is modelled by explanatory variables $a$ and $b$. McIntosh~\cite{mcintosh2015emse} used the following model for Qt:

\begin{align*}
&log(defects + 1) \sim rcs(size, 5) + rcs(all~authors, 5) \\
&\quad + complexity + churn + rcs(change~entropy,3) \\
&\quad + rcs(changes~w/o~discussion,3) \\
&\quad + rcs(self-approved~changes, 5) \\
&\quad + rcs(typcal~discussion~length, 5) \\
&\quad + rcs(typical~reviewer~expertise, 5) + rcs(lacking~subject~matter~expertise,5)
\end{align*}

Our Qt model is the following: 
\begin{align*}
&log(defects + 1) \sim rcs(size, 5) + rcs(prior~defects, 5) \\
&\quad + rcs(churn, 3) + rcs(change entropy,3) \\
&\quad + rcs(all~authors, 5) + rcs(changes~w/o~discussion, 5) \\
&\quad + self-approved~changes + typcal~discussion~length \\
&\quad + typcal~review~window +  rcs(rushed~reviews,3) \\
&\quad + rcs(lacking~subject~matter~expertise,3) + typical~reviewer~expertise
\end{align*}

Our Chrome model is the following:
\begin{align*}
&log(defects + 1) \sim rcs(size, 5) + rcs(prior~defects, 5) \\
&\quad + complexity + rcs(churn, 3) + rcs(change~entropy,3) \\
&\quad + rcs(all~authors, 5) + rcs(self-approved~changes, 5) \\
&\quad + typical~discussion~length + rushed~reviews\\
&\quad + typcal~review~window + rcs(lacking~subject~matter~expertise,3) \\
&\quad + typical~reviewer~expertise
\end{align*}

Restricted Cubic Splines are represented by the rcs function in the formula where the first argument is the predictor and the second argument is the number of knots (the amount of nonlinearity) permitted. 
We fit OLS model using formulas from above and calculate adjusted $R^{2}$ to assess goodness of fit. Tables \ref{results_tbl} and \ref{results_tbl_chrome} summarize the results and contain the original results from McIntosh \etal~\cite{mcintosh2015emse}.
In summary, the response variable is modeled via linear and non-liner dependencies approximated via cubic splines.

\begin{table}
\centering
 \caption{Post-release defects prediction model for Qt. Original and replication study results. Non-linear models do not outperform linear models. While statistically significant, review measures are unstable and have little predictive power compared to traditional measures.}
\label{results_tbl}
\begin{adjustbox}{angle=-90}
\begin{tabular}[\textwidth]{|l|c|c|c|c|c|c|c|c|c|}
\hline
\multicolumn{2}{|l|}{Release}          & \multicolumn{2}{c|}{5.0 McIntosh} & \multicolumn{2}{c|}{5.0}  & \multicolumn{2}{c|}{5.1 McIntosh} & \multicolumn{2}{c|}{5.1} \\ \hline

\multicolumn{2}{|l|}{Nonlinear model adjusted $R^{2}$}               & \multicolumn{2}{c|}{0.69} & \multicolumn{2}{c|}{0.62} & \multicolumn{2}{c|}{0.46} & \multicolumn{2}{c|}{0.66}   \\ 

\multicolumn{2}{|l|}{Linear model adjusted $R^{2}$} & \multicolumn{2}{c|}{} & \multicolumn{2}{c|}{0.61} & \multicolumn{2}{c|}{} & \multicolumn{2}{c|}{0.63}   \\ 

\multicolumn{2}{|l|}{\textbf{Nonlinear model w/o codereview variables adjusted $R^{2}$}}   & \multicolumn{2}{c|}{} & \multicolumn{2}{c|}{\textbf{0.61}} & \multicolumn{2}{c|}{} & \multicolumn{2}{c|}{\textbf{0.64}}  \\ \hline

\multicolumn{2}{|l|}{Overall D.F.}     & \multicolumn{2}{c|}{80}  & \multicolumn{2}{c|}{81} & \multicolumn{2}{c|}{78}  & \multicolumn{2}{c|}{81}   \\ 
\multicolumn{2}{|l|}{Allocated D.F.}   & \multicolumn{2}{c|}{22}  & \multicolumn{2}{c|}{22}   & \multicolumn{2}{c|}{24}  & \multicolumn{2}{c|}{22}  \\ \hline
     \multicolumn{2}{|l|}{}               & Overall    & Nonlinear   & Overall    & Nonlinear & Overall    & Nonlinear   & Overall    & Nonlinear  \\ \hline
Size                   & D.F.              & 4          & 3           & 4          & 3              & 2          & 1           & 4          & 3                     \\
\multicolumn{1}{|l|}{} & $\chi^{2}$        &$110^{***}$  & $76^{***}$  &$60^{***}$  &$14^{**}$ &$10^{**}$  & $5^{*}$  &$47^{***}$  &$35^{***}$   \\ \hline
     
Complexity             & D.F.              & 1          & na          & $\dagger$           & $\dagger$         & 1          & na          & $\dagger$          & $\dagger$        \\ 
\multicolumn{1}{|l|}{} & $\chi^{2}$        &$1^{\circ}$ & na          & $\dagger$   &$\dagger$         &$<1^{\circ}$ & na          & $\dagger$  & $\dagger$  \\ \hline
     
Prior defects          & D.F.              & $\ddagger$ & $\ddagger$  & 2          & 1         & 2          & 1           & 3          & 2   \\ 
\multicolumn{1}{|l|}{} & $\chi^{2}$        & $\ddagger$ & $\ddagger$  & $48^{***}$& $45^{***}$&$9^{*}$   & $<1^{\circ}$  & $90^{***}$&  $77^{***}$ \\ \hline
     
Churn                  & D.F.              & 1          & na          & 2          & 1         & 1          & na          & 2          & 1  \\ 
\multicolumn{1}{|l|}{} & $\chi^{2}$        & $1^{\circ}$&  na & $15^{***}$ & $7^{***}$ &$<1^{\circ}$ & na   & $5^{\circ}$ & $3^{\circ}$ \\ \hline
     
Change entropy         & D.F.              & 2          & 1           & 1          & na        & 2          & 1           & 2          & 1  \\ 
\multicolumn{1}{|l|}{} & $\chi^{2}$        &$8^{*}$     & $7^{**}$     &$<1^{\circ}$& na        &$6^{*}$     & $6^{*}$     &$3^{\circ}$& $2^{\circ}$  \\ \hline
         
All authors            & D.F.              & $\ddagger$    & $\ddagger$    & 3 & 2 & 2          & 1           & 2 & 1  \\ 
\multicolumn{1}{|l|}{} & $\chi^{2}$        & $\ddagger$& $\ddagger$ & $274^{***}$  & $63^{***}$ & $30^{***}$& $15^{***}$  & $193^{***}$    & $13^{***}$   \\ \hline
         
Minor authors               & D.F.              & $\dagger$ & $\dagger$  & $\ddagger$ & $\ddagger$ & 1 & na  & $\ddagger$ & $\ddagger$ \\ 
\multicolumn{1}{|l|}{} & $\chi^{2}$        & $\dagger$ & $\dagger$  & $\ddagger$ & $\ddagger$ & $2^{\circ}$ &na  & $\ddagger$ & $\ddagger$ \\ \hline
         
Major authors          & D.F.              & $\dagger$  & $\dagger$   & $\dagger$  & $\dagger$  & $\dagger$  & $\dagger$   & $\dagger$  & $\dagger$  \\ 
\multicolumn{1}{|l|}{} & $\chi^{2}$        & $\dagger$  & $\dagger$   & $\dagger$  & $\dagger$  & $\dagger$  & $\dagger$   & $\dagger$  & $\dagger$ \\ \hline
         
Author ownership       & D.F.              &  $\dagger$ & $\dagger$   &$\dagger$         & $\dagger$         &  $\dagger$ & $\dagger$   & $\dagger$          & $\dagger$  \\
\multicolumn{1}{|l|}{} & $\chi^{2}$        &  $\dagger$ & $\dagger$   & $\dagger$ & $\dagger$         &  $\dagger$ & $\dagger$   & $\dagger$ & $\dagger$ \\ 
\hline
\hline
Self-approved          & D.F.              & 2          & 1           &1          & na         & 1          & na           & 1         & na \\ 
\multicolumn{1}{|l|}{} & $\chi^{2}$        & $22^{***}$& $1^{\circ}$  & $7^{*}$    & $2^{\circ}$& $<1^{\circ}$ & na & $<1^{\circ}$    & na  \\ \hline
     
Rushed reviews         & D.F.              & $\dagger$  & $\dagger$   & 2 & 1 & 2                   & 1             &2  & 1 \\ 
\multicolumn{1}{|l|}{} & $\chi^{2}$        & $\dagger$  & $\dagger$   & $4^{\circ}$ &  $<1^{\circ}$  & $48^{***}$ & $23^{***}$   & $3^{\circ}$  & $<1^{\circ}$  \\ \hline
     
Changes w/o disc.      & D.F.              &  2         & 1           & 2          & 1         &  2         & 1           & 2          & 1 \\ 
\multicolumn{1}{|l|}{} & $\chi^{2}$        &$6^{\circ}$     & $4^{*}$ & $18^{**}$& $3^{\circ}$         &$3^{\circ}$ & $1^{\circ}$ & $36^{***}$& $29^{***}$ \\ \hline
     
Typical review window  & D.F.              & $\dagger$  &  $\dagger$  & 1  & na  & $\dagger$  &  $\dagger$  & 1  & na \\ 
\multicolumn{1}{|l|}{} & $\chi^{2}$        & $\dagger$  &  $\dagger$  & $<1^{\circ}$  & na  & $\dagger$  &  $\dagger$  & $<1^{\circ}$ & na  \\ \hline
Typical disc. length   & D.F.              & 4          & 3           & 1          & na         & 2          & 1           & 1          & na  \\ 
\multicolumn{1}{|l|}{} & $\chi^{2}$        &$26^{***}$  & $24^{***}$   & $<1^{\circ}$  & na  &$32^{***}$  & $21^{**}$   & $3^{\circ}$ & na \\ \hline
\hline
Lacking subject matter expertise   & D.F.              & 2          & 1           & 2          & 1        & 4         & 3           & 1          & na  \\ 
\multicolumn{1}{|l|}{} & $\chi^{2}$        &$80^{***}$  & $70^{***}$   & $33^{***}$  & $3^{\circ}$ &$34^{***}$  & $22^{**}$   & $<1^{\circ}$ & na \\ \hline
Typical reviewer expertise  & D.F.              & 4          & 3           & 1          & na          & 2          & 1           & 1          & na \\ 
\multicolumn{1}{|l|}{} & $\chi^{2}$        &$26^{***}$  & $24^{***}$   & $<1^{\circ}$  & na  &$32^{***}$  & $21^{**}$   & $7^{**}$  & na \\ \hline
\multicolumn{10}{l}{Discarded during: $\dagger$ - Removed during correlation analysis; $\ddagger$ - Removed during redundancy analysis}  \\ 
\multicolumn{10}{l}{Statistical significance: $***  \rho < 0.001$; $**  \rho < 0.01$; $*  \rho < 0.05$; $\circ  \rho >= 0.05$}   \\ 
\multicolumn{10}{l}{Other: na - not used} 
\end{tabular}
\end{adjustbox}
\end{table}        

\begin{table}
\centering
\caption{Post-release defects prediction model for Chrome. Non-linear models do not outperform linear models. While statistically significant, review measures are unstable and have little predictive power compared to traditional measures.}
\label{results_tbl_chrome}
\begin{adjustbox}{angle=-90}
\begin{tabular}{|l|l|c|c|c|c|c|c|c|c|c|c|c|c|}
\hline
\multicolumn{2}{|l|}{Release}  & \multicolumn{2}{c|}{39}  & \multicolumn{2}{c|}{40} & \multicolumn{2}{c|}{41} & \multicolumn{2}{c|}{42}  & \multicolumn{2}{c|}{43} & \multicolumn{2}{c|}{44} \\ \hline

\multicolumn{2}{|l|}{Nonlinear model adjusted $R^{2}$}  & \multicolumn{2}{c|}{0.61}   & \multicolumn{2}{c|}{0.58} & \multicolumn{2}{c|}{0.59} & \multicolumn{2}{c|}{0.59} & \multicolumn{2}{c|}{0.51} & \multicolumn{2}{c|}{0.53}   \\ 

\multicolumn{2}{|l|}{Linear model adjusted $R^{2}$} & \multicolumn{2}{c|}{0.59} & \multicolumn{2}{c|}{0.54} & \multicolumn{2}{c|}{0.56} & \multicolumn{2}{c|}{0.57}   & \multicolumn{2}{c|}{0.50} & \multicolumn{2}{c|}{0.49}  \\ 

\multicolumn{2}{|l|}{\textbf{w/o codereview variables adjusted $R^{2}$}}  & \multicolumn{2}{c|}{\textbf{0.60}}   & \multicolumn{2}{c|}{\textbf{0.58}} & \multicolumn{2}{c|}{\textbf{0.59}} & \multicolumn{2}{c|}{\textbf{0.59}} & \multicolumn{2}{c|}{\textbf{0.49}} & \multicolumn{2}{c|}{\textbf{0.53}}   \\ 
\hline

\multicolumn{2}{|l|}{Overall D.F.}     & \multicolumn{2}{c|}{62}  & \multicolumn{2}{c|}{87} & \multicolumn{2}{c|}{90}  & \multicolumn{2}{c|}{84}   & \multicolumn{2}{c|}{83}  & \multicolumn{2}{c|}{80}  \\ 
\multicolumn{2}{|l|}{Allocated D.F.}   & \multicolumn{2}{c|}{26}  & \multicolumn{2}{c|}{21}   & \multicolumn{2}{c|}{23}  & \multicolumn{2}{c|}{20}   & \multicolumn{2}{c|}{19}  & \multicolumn{2}{c|}{20}  \\ \hline
     \multicolumn{2}{|l|}{}               & Overall    & Nonlinear   & Overall    & Nonlinear & Overall    & Nonlinear   & Overall    & Nonlinear   & Overall    & Nonlinear   & Overall  & Nonlinear   \\ \hline
Size                   & D.F.              & 4          & 3           & 2          & 1              & 4          & 3           & 2         & 1                    & 2          & 1           & 2        & 1  \\
\multicolumn{1}{|l|}{} & $\chi^{2}$        &$28^{***}$ & $3^{\circ}$  &$1^{\circ}$  &$1^{\circ}$ &$10^{*}$  & $<1^{\circ}$  &$1^{\circ}$  &$<1^{\circ}$   & $3^{\circ}$& $2^{\circ}$ & $8^{*}$      & $6^{*}$    \\ \hline
     
Complexity             & D.F.              & 1       &na            & 1          & na        & 1          & na          & 1          & na          & 1          & na          & 1        &na            \\ 
\multicolumn{1}{|l|}{} & $\chi^{2}$        & $<1^{\circ}$&  na         & $3^{\circ}$  &na         &$<1^{\circ}$ & na         & $<1^{\circ}$  &na           & $1^{\circ}$& na          & $<1^{\circ}$        & na    \\ \hline
     
Prior defects          & D.F.              & 4 & 3  & 2          & 1         & 4          & 3          & 4         & 3  & 4          & 3           & 2        & 1           \\ 
\multicolumn{1}{|l|}{} & $\chi^{2}$        & $43^{***}$ & $42^{***}$  & $37^{***}$ & $34^{***}$&$74^{***}$   & $71^{***}$  & $61^{***}$&$55^{***}$  & $21^{***}$ & $20^{***}$  & $3^{\circ}$        & $3^{\circ}$    \\ \hline
     
Churn                  & D.F.              & 2          & 1         & 2          & 1         & 2          & 1          & 2          & 1  & 2          & 1           & 2        & 1          \\ 
\multicolumn{1}{|l|}{} & $\chi^{2}$        & $14^{**}$& $8^{*}$   & $23^{***}$ & $22^{***}$ &$31^{***}$ & $30^{***}$    & $26^{***}$ & $24^{***}$  & $<1^{\circ}$ & $<1^{\circ}$  & $11^{**}$        & $2^{\circ}$   \\ \hline
     
Change entropy         & D.F.              & 2          & 1           & 1          & na        & 1          & na          & 1          & na   & 1          & na         & 1        & na           \\ 
\multicolumn{1}{|l|}{} & $\chi^{2}$        &$<1^{\circ}$     &$<1^{\circ}$     &$<1^{\circ}$& na        &$<1^{\circ}$     & na     &$<1^{\circ}$& na  &$3^{*}$& na          & $<1^{\circ}$        & na    \\ \hline
         
All authors            & D.F.              & 3          & 2           & 4 & 3  & 4  & 3           & 3 & 2   & 3          & 2           & 3        & 2           \\ 
\multicolumn{1}{|l|}{} & $\chi^{2}$        &$49^{***}$& $2^{\circ}$ & $43^{***}$ & $8^{*}$& $42^{***}$& $2^{\circ}$  & $33^{***}$    & $<1^{\circ}$  & $51^{***}$ & $4^{\circ}$    & $24^{***}$       & $1^{\circ}$    \\ \hline
         
Minor authors               & D.F.              & $\ddagger$ & $\ddagger$  & $\ddagger$ & $\ddagger$ & $\ddagger$ & $\ddagger$  & $\ddagger$ & $\ddagger$ & $\ddagger$ & $\ddagger$  & $\ddagger$        & $\ddagger$           \\ 
\multicolumn{1}{|l|}{} & $\chi^{2}$        & $\ddagger$ & $\ddagger$  & $\ddagger$ & $\ddagger$ & $\ddagger$ &$\ddagger$  &$\ddagger$ &$\ddagger$&$\ddagger$ & $\ddagger$  & $\ddagger$        & $\ddagger$    \\ \hline
         
Major authors          & D.F.              & $\dagger$ & $\dagger$   & $\dagger$  & $\dagger$  & $\dagger$  & $\dagger$  & $\dagger$  & $\dagger$ & $\dagger$  & $\dagger$   & $\dagger$        & $\dagger$        \\ 
\multicolumn{1}{|l|}{} & $\chi^{2}$        & $\dagger$  & $\dagger$   & $\dagger$  & $\dagger$  & $\dagger$  & $\dagger$   & $\dagger$  & $\dagger$ & $\dagger$ & $\dagger$   & $\dagger$        & $\dagger$   \\ \hline          
Author ownership       & D.F.              &  $\dagger$ & $\dagger$   & $\dagger$    & $\dagger$   & $\dagger$ & $\dagger$   & $\dagger$  & $\dagger$ &$\dagger$  & $\dagger$   & $\dagger$  & $\dagger$           \\
\multicolumn{1}{|l|}{} & $\chi^{2}$        & $\dagger$ & $\dagger$  & $\dagger$ & $\dagger$&$\dagger$ &$\dagger$  & $\dagger$ & $\dagger$ & $\dagger$ & $\dagger$  & $\dagger$       & $\dagger$    \\ 
\hline
\hline
Self-approved          & D.F.              & 4          & 3           & 4          & 3          & 4          & 3           & 4        & 3 & 4          & 3           & 4        & 3           \\ 
\multicolumn{1}{|l|}{} & $\chi^{2}$        & $8^{*}$& $7^{\circ}$ & $12^{*}$  & $5^{\circ}$& $2^{\circ}$& $2^{\circ}$ & $3^{\circ}$    & $2^{\circ}$ & $13^{**}$  & $9^{**}$     & $9^{*}$        & $3^{\circ}$    \\ \hline
     
Rushed reviews         & D.F.              & 1  & na   & 1  & na  & 1                  & na             & 1  & na & 1        & na         & 1        & na           \\ 
\multicolumn{1}{|l|}{} & $\chi^{2}$        & $1^{\circ}$  & na & $19^{***}$ & na  &$2^{\circ}$ & na   & $14^{**}$  & na & $6^{*}$ & na          & $<1^{\circ}$        & na    \\ \hline
     
Changes w/o disc.      & D.F.              &  $\dagger$   & $\dagger$  & $\dagger$    & $\dagger$ & $\dagger$ & $\dagger$   & $\dagger$& $\dagger$ & $\dagger$  & $\dagger$   & $\dagger$        & $\dagger$           \\ 
\multicolumn{1}{|l|}{} & $\chi^{2}$        &$\dagger$    &$\dagger$ & $\dagger$& $\dagger$         &$\dagger$ & $\dagger$ & $\dagger$& $\dagger$ & $\dagger$  & $\dagger$   & $\dagger$        & $\dagger$    \\ \hline
     
Typical review window  & D.F.              & 1  &  na  & 1 & na  & 1  &  na  & 1  & na & 1          & na        & 1       & na           \\ 
\multicolumn{1}{|l|}{} & $\chi^{2}$        & $1^{\circ}$  &  na  & $1^{\circ}$& na   & $<1^{\circ}$  &  na  & $<1^{\circ}$  & na & $<1^{\circ}$& na          & $<1^{\circ}$        & na   \\ \hline
Typical disc. length   & D.F.              & 1         & na          & 1          & na          & 1          & na           & 1          & na & 1          & na          & 1 & na      \\ 
\multicolumn{1}{|l|}{} & $\chi^{2}$        &$<1^{\circ}$ & na   & $<1^{\circ}$& na   &$<1^{\circ}$  & na   & $<1^{\circ}$  & na &$<1^{\circ}$& na         & $<1^{\circ}$        & na    \\ \hline
\hline
Lacking subject matter expertise   & D.F.              &2         & 1          & 2          & 1          &2          & 1           & 2          & 1 & 1          & na          & 2 & 1      \\ 
\multicolumn{1}{|l|}{} & $\chi^{2}$        &$7^{*}$ & $3^{\circ}$   & $3^{\circ}$& $<1^{\circ}$   &$32^{***}$  & $7^{***}$   & $20^{***}$  & $5^{*}$ &$7^{**}$& na         & $19^{***}$        &$10^{**}$    \\ \hline
Typical reviewer expertise   & D.F.              & 1         & na          & 1          & na          & 1          & na           & 1          & na & 1          & na          & 1 & na      \\ 
\multicolumn{1}{|l|}{} & $\chi^{2}$        &$<1^{\circ}$ & na   & $<1^{\circ}$& na   &$<1^{\circ}$  & na   & $<1^{\circ}$  & na &$7^{*}$& na         & $10^{*}$        & na    \\ \hline
\multicolumn{14}{l}{Discarded during: $\dagger$ - Removed during correlation analysis; $\ddagger$ - Removed during redundancy analysis}  \\ 
\multicolumn{14}{l}{Statistical significance: $***  \rho < 0.001$; $**  \rho < 0.01$; $*  \rho < 0.05$; $\circ  \rho >= 0.05$}   \\ 
\multicolumn{14}{l}{Other: na - not used} 
\end{tabular}
\end{adjustbox}
\end{table}   

\subsection{Model results and model comparisons}

In this section we compare our replication results with those from McIntosh \etal~\cite{mcintosh2015emse} original study and new results from Chrome. We highlight differences and discuss their possible causes.

\subsection{Comparing linear and non-linear models}

\begin{figure}
 \centering
 \includegraphics[width=0.7\textwidth]{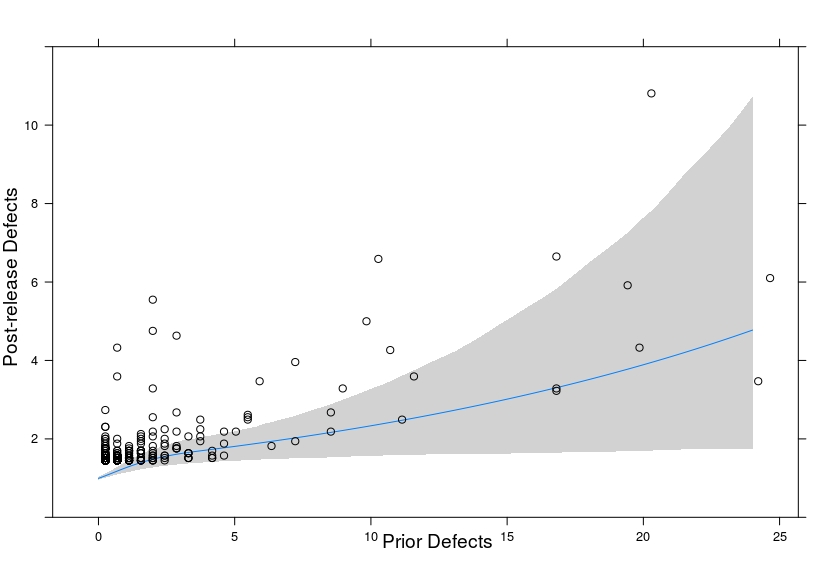}

 \caption{A wide margin of error for nonlinear predictions, for example, prior defects in Qt 5.0. Nonlinear models are unnecessary, see Tables~\ref{results_tbl} and \ref{results_tbl_chrome}.}
 \label{nonlinear-priordefects-qt50}
\end{figure}

To illustrate the nonlinear effect we select an independent variable
with the highest potential of nonlinearity from the model and
calculate predicted number of post-release defects as the function
of this variable, using \emph{Predict} function from R rms
package. The rest of the variables are fixed at their median
values. As an illustration, we choose \emph{prior defects} for Qt
because it had the highest Spearman squared value among independent variables participating in the model. We then plot the results in Figure \ref{nonlinear-priordefects-qt50}. Although the shape of the plot may suggest some nonlinearity, the grey funnel, which is the error margin, is too wide to claim with confidence that these variables have a nonlinear relation with the response variable. The goodness of fit $R^2$ also shows that nonlinear models do not yield better results than regular linear models. Discussion with McIntosh revealed that the text of the original paper was ambiguous and they only log transformed the dependent variable. We find that a log transformation of skewed independent variables provides a reasonable model without adding the complexity of a non-linear model.

\subsection{Models with and without review measures}

The results show that although many of the code review measures are statistically significant, they usually tend to have lower values of Wald $\chi^{2}$ test than other measures, suggesting their lower contribution to explanatory power of the model. Even the most prominent measures, like \emph{typical discussion length} and \emph{rushed reviews} are repeatedly outperformed by measures like \emph{size}, \emph{prior defects}, and \emph{all authors}.
As a further investigation, we fit a model {\it without} review measures and record the values of adjusted $R^{2}$ (shown in bold in the section of $R^{2}$ values in the Tables \ref{results_tbl}, \ref{results_tbl_chrome}). The decrease in the values of adjusted $R^{2}$ in both datasets is minimal, meaning that overall contribution of the review measures to the explanatory power of the model is limited. 
In addition to low contribution to the model the performance of
review variables is inconsistent between datasets and releases. For
instance, in McIntosh \etal \emph{rushed reviews} was discarded from
the model in Qt 5.0 during the correlation analysis, however, in Qt
5.1 this measure is one of the strongest predictors of post-release defects. The \emph{typical discussion length} is one of the most influential variables for both Qt releases in both studies, but in Chrome dataset the contribution of this variable is insignificant. Discussing our results with McIntosh, he stated that he did not believe that review measures could dominate traditional measures. Our answer to RQ1, Replication, is summarized below.

\vspace{+5mm}
\fbox{ 
\begin{minipage}[b]{.9\textwidth}
{\bf Conclusion 1:} {\it The review measures contributed little to the performance of the model, with the $R^{2}$ remaining almost unchanged from the model that included only the traditional predictors, such as the number of prior-defects, size, and authors.}
 \end{minipage}
}

\subsection{Impact of individual variables}

\paragraph{\textbf{Size of component}} is a well-known predictor in empirical software studies. McIntosh \etal show that in the Qt project \emph{size} provides significant contribution to the explanatory power of the model. Our result is similar for Qt dataset. However, in Chrome dataset the contribution of the \emph{size} measure is quite small (Table \ref{results_tbl_chrome}). 

\paragraph{\textbf{Prior defects and all authors}} have been shown to be good predictors of future defects \cite{graves2000predicting,bird2011don}. 
McIntosh~\etal discard \emph{prior defects} in Qt 5.0 due to
redundancy. In our study, the redundancy analysis on the Qt 5.0
dataset does not indicate that \emph{prior defects} are
redundant and, on the contrary, is statistically significant. For the Qt 5.1 release, both McIntosh \etal and our model
keep \emph{prior defects} but find it to be a poor predictor. The
\emph{all authors} measure is redundant in Qt 5.0 release in our
study contrary to the McIntosh~\etal. For Qt 5.1 the \emph{all
  authors} is the most influential predictor in the model. This
result is replicated in both studies. In Chrome dataset these two
variables are repeatedly found to be the most influential variables
of the model. A possible explanation for this inconsistency could be
that these two variables share a common cause. Defects are not
always fixed by the owner of the module, especially in big
teams. That means that more developers are touching the file, and
the more developers modifying a file the higher the risk of the
future defects. Intuitively, the growth in these two measures should
be related, but our correlation and redundancy analysis fails to
find this. These inconsistent results suggest that traditional
variable selection techniques are not capable of coping with the
high correlations in our datsets  and indicate the need for
a different approach that can deal with complex interactions between
variables.  

\paragraph{\textbf{Review measures}}. The important measures in the
Qt dataset are similar to what McIntosh \etal found. The
\emph{self-approved changes} has low impact on post-release
defects. The \emph{rushed reviews} variable was discarded in 5.0
release, but in 5.1 it appears as one of the most influential
variables. The \emph{typical discussion length} variable has
moderate to strong influence in both releases.  For the Chrome
dataset the review measures are not statistically significant in
most cases. When they are, such as in \emph{lacking subject matter
  expertise} the result is inconsistent across releases. The overall
performance of review variables is inconsistent in both studies suggesting the following conclusion to RQ 2, Differentiated-external replication. 

%\textbf{Author ownership} is another measure which was observed in the past to affect the overall quality of code~\cite{bird2011don}. Our results agree with this observation, \emph{author ownership} is statistically significant and has relatively high of Wald $\chi^{2}$ (25, 33) in both considered Qt releases. In contrast, McIntosh \etal miss this results because they discarded the variable as it was highly correlated. We did not find that it was highly correlated during our analysis and did not remove it.

\vspace{+5mm}
\fbox{ 
\begin{minipage}[b]{.9\textwidth}
{\bf Conclusion 2:} {\it The inconsistent performance across projects and releases of strong predictors, like \emph{all authors}, \emph{prior defects} and others, suggests a possible issue with the traditional variable selection approach and indicates the need for an approach that is capable of dealing with a more complex inter-variable relations.}
 \end{minipage}
}

%\begin{figure}
       %\centering
   %\includegraphics[width=0.5\textwidth]{images/cluster_chrome40}
   %\caption{Hierarchical Correlation Analysis for Chrome 40}
   %\label{FigCorChrome40}
%\end{figure}

\section{Bayesian Networks Models} \label{secBN}

To address the concern of the
unexpected absence of the relationship between code review measures and
post-release defects demonstrated in the previous models, and the lack of
reproducibility due to the subjectivity in variable selection
approaches that are necessary in a traditional model, such as the one used
in Section~\ref{secReplication}, we use Bayesian Networks (BN) as an alternative 
modeling approach. Our goal in using the BN model is not to create the best predictive 
model for post-release defects. Instead, we focus on understanding the complex 
interaction between the variables described by the data, and determining which variables 
directly impact the number of post-release defects in such a generative 
model.\footnote{A generative model
 specifies a joint probability distribution over all observed
 variables, whereas a discriminative model provides a model only
 for the target variable(s) conditional on the predictor
 variables. Thus, while a discriminative model allows only sampling
 of the target variables conditional on the predictors, a
 generative model can be used, for example, to simulate
 (\ie generate) values of any variable in the model, and
 consequently, to gain an understanding of the underlying mechanics
 of a system, generative models are essential.}
 %We introduce provide the background on Bayesian Networks, describe the model creation, and discuss the results.

\subsection{Background: Bayesian Network}
\label{BN}
\label{vm_construction} 

Bayesian Network models have several advantages over
regression models. To be precise, regression analysis is a very
simple BN where there is one directed link from each independent variable
to the dependent variable. BNs, therefore, can help with 
multicollinearity by establishing the relationships among independent variables. In the process of BN
construction we can control the number of edges (relations) by
specifying a connection strength threshold. Once the
Bayesian Network is constructed we can use the graphical
representation to learn about less obvious interactions among
variables and infer how the injection of specific facts affects
variables of interest. We use BN to investigate the lack of
consistency in the replication in the previous sections.

%we already said this:
%A Probabilistic Graphical Model (PGM) is a mathematical way to encode the probabilistic relationships, such as conditional independence and dependence, among the random variables. The nodes of a PGM are the random variables and the edges represent the probabilistic dependence. A PGM tells us how we believe the variables are related in our probabilistic generative model. Such a model, with all variables of importance, can provide an understanding of the underlying mechanics, and help to discover if the review measures are related to post-release defects, and to understand which variables play a major role in understanding software quality.

%\todo{Audris: why do we talk about two varieties? Do we use both? I found this confusing already, perhaps you can limit this to what is actually used?} PGMs come in two varieties, Bayesian networks (which we use in this paper) representing directed acyclic graphs (DAGs) and Markov random fields representing undirected graphs. They differ in the set of independencies they can encode and the factorization of the distribution that they induce~\cite{koller2009probabilistic}.
%

One important concept related to the BNs is the concept of \emph{Markov 
Blanket}~\cite{pearl2014probabilistic}. The Markov Blanket for a node  
in a Bayesian Network is the set of nodes composed of its parents, 
its children, and its children's other parents (co-parents). 
The Markov blanket of a node contains all the variables that 
shield the node from the rest of the network \ie for a node $A$, 
its Markov Blanket $MB_{A}$, and a node $B: B \neq A, B \notin MB_{A}$, 
we have the property that: $$ Pr(A | MB_{A}, B) = Pr(A|MB_{A}) $$
This means that the Markov blanket of a node is the only knowledge 
needed to predict the behavior of that node. 

We will construct a BN model without assuming any domain expertise, edges, or assumptions about prior data distributions. Instead we will use minimal a-priori model that focuses on the search for the best Bayesian graphical representation for the dataset (\ie structure learning using hill climbing).
%There are two primary ways of constructing BN models. In the first approach the graph represents dependencies obtained from domain experts.  The graph may include prior distributions about the parameters of the overall model. The data is then used to calculate the posterior distribution and to make inference. The second approach puts minimal a-priori assumptions about the model and focuses on the search for the best graphical representation for a given dataset (structure learning). This is an NP-hard problem~\cite{chickering1996learning}, but a number of different heuristic structure learning algorithms are available. 
%Just discuss this in limitations.
%For simplicity, we will assume no latent or hidden variables, \ie, we assume that the set of variables we use fully describes our problem domain (discussed further in Section~\ref{sec:limitations}).

%\subsection{Model construction} \label{vm_construction} 
%
%Here we describe the procedures used to create our model and the 
%reasoning behind them. 

Despite the promises of BNs, they tend to be quite sensitive to data, and operational data is often
problematic~\cite{M14,zmz15}. Careful preprocessing is needed to ensure a reliable and reproducible
result. We next discuss discretization of variables and  structure learning with hill climbing approaches used to address these concerns.

\subsubsection{Discretization}
\label{secDiscretizationMethod}

For our regression model, we found that all our variables have a 
long-tailed distribution that could not be
corrected even by a log-transformation. Since BN
structure learning methods for continuous data require a normal
distribution, we discretize the data, as is often done with prediction that involves classifiers~\cite{bnppt}. 
Discretizing variables while preserving relationships among them is
an NP-hard problem~\cite{chlebus1998finding}, 
but several heuristics exist. The commonly used supervised methods optimize
discretization to improve explanatory power for 
a single response variable, such as, Chi-square, or
MDLP. However, these are not suitable for a BN structure search, because we do
not know {\em a-priori} which variables will be responses (have arrows
pointing to them) and which will be independent (have no incoming
arrows). While some research on multidimensional
discretization methods exist~\cite{perez2006supervised},
we are not aware of any such method that has a robust implementation in a statistical package.
We, therefore, use unsupervised discretization methods. The added benefit is that
the discretization was totally response-variable agnostic unlike
the commonly used supervised discretization methods, which 
prevents any bias towards specific fit that may accompany supervised
methods. 

Following the recommendations from Garcia \etal's~\cite{garcia2013survey} survey, we use the Equal Frequency discretization method and the implementation in the 
\emph{arules} package. The \textit{defects} node was discretized to a binary no-defect/defect variable, because around 73\% of the directories have no defects, therefore it makes sense to just predict whether or not there will be a defect for our dataset. Two levels were also assigned to minor authors, rushed reviews, typical review window, and lacking subject matter expertise because more than 50\% of entries were zero. Based on the data distributions for the remaining variables three levels were deemed appropriate. We present the distribution of the variables in our replication package~\cite{ReplicationPackage} as additional evidence for the choice of our discretization levels.

\subsubsection{BN Structure Search: Hill Climbing}

To learn the BN structure from our data, we chose a well-performing and widely
used~\cite{dey2020deriving} Hill-Climbing (HC) algorithm from the \emph{bnlearn} R package. 
%We chose this package because prior work has found it to be the best performing modeling technique in terms of the accuracy of the final model as well as runtime in an extensive simulation study described in. 
%\todo{audris, where in this paper or in the related work?}
%, where this technique was found to produce no more than 1 missing/extra/reversed edge in 86\% of the cases. 
%We used the implementation of the algorithm as available in 
%the \emph{bnlearn} R package. Implementation details for this algorithm are outside the
%scope of this paper. 
This HC algorithm attempts to maximize the network score with several scoring functions 
available in the \emph{bnlearn} package: \eg BIC, AIC, BDE.  A detailed 
study examining how well different scores performed 
concluded that in general all scores perform similarly and for large data
sets Bayesian scores are more suitable~\cite{carvalho2009scoring}. Since our dataset is not particularly large, at least for 
the individual releases, we decided not to use Bayesian scores \eg BDE, instead we 
chose to focus on the information theoretic scores \eg AIC, BIC. We finally selected the BIC score 
because it is more appropriate for constructing explanatory models, while AIC is better 
suited for building predictive models~\cite{sober2002instrumentalism,shmueli2010explain}.

Hill-Climbing has the known limitation of finding a local
maxima, and there are several enhanced versions of the algorithm that deal
with this shortcoming. The R implementation provides parameters for the
number of random restarts and perturbations as tuning parameters to deal
with this problem. However, these parameters can make the results noisy,
with different settings inducing slightly different networks. To mitigate
this effect, we use the non-parametric bootstrap model averaging method,
which provides confidence levels for both the existence of an edge and its
direction~\cite{friedman1999data}. This enables us to select a model based on
a confidence threshold. Friedman \etal \cite{friedman1999data} argued that
the threshold is domain specific and needs to be determined for each domain. 
To identify a suitable threshold, we performed a simulation study, by
generating a simulated dataset for the same number of nodes. The result of
the simulation
showed that a threshold of $0.65$ was suitable to accurately recover the
original structure. We also investigated alternative thresholds to assess the
stability of the results as described in Section~\ref{sec:limitations}.

Finally, due to the HC algorithm not being a deterministic one, we repeated 
the process of generating a model 100 times, according to the recommendation
by Arcuri and Briand~\cite{arcuri2011practical} and generated our final
model based on the averaged result of these 100 runs. 
%We, however, did not
%repeat the process 1000 times, as~\cite{arcuri2011practical} recommended,
%because that would take around 3 months to run.

\subsubsection{Combining Data for Qt and Chrome Releases}
We created new datasets by combining the data for all the Qt releases, the Chrome releases, and also combined the data for all releases for the two projects, which gave us three new datasets. Combining the datasets makes our final model more robust and the result is arguably more generalizable, because of having more training examples and not being prone to overfit to the specific characteristics of one particular release. 

\subsubsection{BN Model Performance Measures}

Model performance can be evaluated using explanatory and predictive performance measures~\cite{shmueli2010explain}. We first create an explanatory model that allows us to understand which software engineering measures have the greatest influence on post-release defects. We use the variance explained, which is the proportion of the log-likelihood score of the model relative to the baseline model which assumes that all the variables are independent. 

We test the predictive power of our BN models for the purpose 
of comparison between similar models and to demonstrate the practical applicability of the models. Training and testing the models with Cross-fold validation is \textit{not} appropriate because we have time ordered data. To ensure that we are not using future data to predict past observations, we trained our model on the earlier 70\% of the data and test it on the subsequent 30\% of the data. We use the \textit{Accuracy} and Cohen's \textit{Kappa} measures as the performance measures for our models. 
%We use Kappa, Cohen's kappa coefficient, because it is considered to be more robust than simple percentage of agreement. 

\subsubsection{CPT: the probability of having a defect given each variable}

To determine the individual impact that each variable has on post-release defects, we create Conditional Probability Tables (CPT) for each variable in the  Markov Blanket that directly influences defects. The CPTs are trained with the \emph{gRain} package in R, using Junction Tree 
belief propagation method (Lauritzen-Spiegelhalter  algorithm~\cite{GRAINgraphical,lauritzen1988local,koller2009probabilistic}) from the BN models. See~\cite{GRAINgraphical} for details about how the CPT tables are calculated. The code for CPT construction is available in our replication package~\cite{ReplicationPackage}.

%%% The CPT calculation is fairly involved, so I don't want to go into the details of it. It involves Moral Graphs and Triangulation - Tapajit 

% A BN is a simple way of specifying a multivariate distribution by combining  (conditional) univariate distributions. For vertex $v$ in a BN with $V$ nodes, the probability distribution for $X_v$, a discrete random vector associated with the vertex $v$, will take the form:$ p(x_v) = \prod_{v \in V} p(x_v| x_{pa(v})$, where $x_v$ denotes the levels of $X_v$,  $pa(v)$ denotes the parents of the vertex $v$, and $p(x_v| x_{pa(v})$ is a function defined on $X_v \times X_{pa(v)}$. This function is non-negative and satisfies that 
% $\sum_{x_v} p(x_v| x*_{pa(v}) = 1$ for each parent configuration $x*_{pa(v)}$ of $x_{pa(v)}$. Hence, $x_{pa(v)}$ becomes the conditional distribution of $X_v$ given $X_{pa(v)}$, which is most commonly specified as a table called a conditional probability table (CPT). Thus, a Bayesian Network can be regarded as a complex stochastic model built up by putting together simple components (conditional probability distributions)~\cite{GRAINgraphical}.

%Finally, setting type = "conditional" gives p(defects |size , finding), i.e., the distribution of the first variable in nodes given the remaining variables listed.

\subsection{Results for BN models}

\begin{table}[t]
\centering
\caption{Variables in Markov Blanket of \textit{defects} for each BN model and the model performance as measured by Variance Explained, Accuracy, and Kappa.}
\label{t:MB_release}
\resizebox{\textwidth}{!}{%
\begin{tabular}{lp{4cm}rrr}
\hline
\textbf{Release} & \textbf{Variables in Markov Blanket of \textit{defects}} & \textbf{Variance Explained} & \textbf{Accuracy} & \textbf{Kappa} \\ \hline
Qt\_50 & priordefects, changesnodisc & 51.4\% & 0.85 & 0.43 \\
Qt\_51 & allauthors, size & 32.9\% & 0.86 & 0.44 \\ \hline
Qt\_combined & priordefects, changesnodisc & 27.2\% & 0.87 & 0.42 \\ \hline
Chrome\_39 & allreviews, priordefects & 19.8\% & 0.75 & 0.5 \\
Chrome\_40 & priordefects & 25.4\% & 0.76 & 0.48 \\
Chrome\_41 & allchangescount, priordefects & 30.7\% & 0.78 & 0.51 \\
Chrome\_42 & priordefects & 32.9\% & 0.77 & 0.46 \\
Chrome\_43 & minorauthors & 25.6\% & 0.76 & 0.42 \\
Chrome\_44 & allchangescount, minorauthors & 36.1\% & 0.8 & 0.48 \\ \hline
Chrome\_combined & minorauthors, priordefects & 19.5\% & 0.76 & 0.41 \\ \hline
All & minorauthors, priordefects, size & 32.7\% & 0.83 & 0.45 \\ \hline
\end{tabular}%
}\end{table}

We created BN models for each release, each project, and for the
entire combined dataset. The results are summarized in
Table~\ref{t:MB_release}. Importantly, the post-release defects
variable, \ie the \emph{defects} node in the Markov Blanket, was
influenced by but did not influence any of the other variables. This
is reassuring, because it is not possible for post-release defects
to influence pre-release measures (going backward in time). Our
models were able to properly account for the fact by correctly
identifying the prior defects in the \emph{priordefects} variable as
having the influence on post-release defects.

Table~\ref{t:MB_release} shows the model performance measures: variance explained, Accuracy, and Kappa. We observe that Kappa values varied between 0.41 and 0.51, which signifies a fair or moderate agreement according to both Landis and Koch~\cite{landis1977application} and Fleiss~\cite{fleiss1981measurement}. Similarly, the accuracy of the models was also observed to be high, between 0.75 and 0.87, and the models had reasonable variance explained, between 19.5\% and 51.4\%.

\subsection{Variables directly affecting post release defects}

\begin{table}[]
\caption{CPT: The conditional probability of post-release defects for each measure. As an example, having one or more minor authors making changes in a component increases the probability of post-release defects by over 70\%.}
\label{t:cpt}
\resizebox{\textwidth}{!}{%
\begin{tabular}{cc||cc}
\hline
\multicolumn{4}{c}{\textbf{For release Qt 50: 12.5\% of observations had defect}} \\\hline
priordefects & probability of defects & changesnodisc & probability of defects \\\hline
0 & 4.8\% & 0 & 5.6\% \\
1 & 7.6\% & 1 & 9.7\% \\
{[}2, 528{]} & 30.6\% & {[}2, 95{]} & 42.3\% \\\hline
\multicolumn{4}{c}{\textbf{For release Qt 51: 17.1 \% of observations had defect}} \\\hline
allauthors & probability of defects & size & probability of defects \\\hline
1 & 6.1\% & {[}0, 78) & 4.1\% \\
2 & 10.5\% & {[}78, 395) & 11.7\% \\
{[}3, 57{]} & 44.1\% & {[}395, 81654{]} & 34.9\% \\\hline
\multicolumn{4}{c}{\textbf{For  Qt - Combined: 14.4 \% of observations had defect}} \\\hline
priordefects & probability of defects & changesnodisc & probability of defects \\\hline
0 & 7.0\% & 0 & 7.3\% \\
1 & 10.4\% & 1 & 14.4\% \\
{[}2, 624{]} & 30.8\% & {[}2, 191{]} & 40.9\% \\\hline
\multicolumn{4}{c}{\textbf{For release Chrome 39: 42.1\% of observations had defect}} \\\hline
priordefects & probability of defects & allreviews & probability of defects \\\hline
{[}0, 5) & 20.7\% & {[}0, 10) & 18.3\% \\
{[}5, 18) & 45.1\% & {[}10, 36) & 39.4\% \\
{[}18, 1588{]} & 64.9\% & {[}36, 1573{]} & 69.8\% \\\hline
\multicolumn{4}{c}{\textbf{For release Chrome 40: 32.2\% of observations had defect}} \\\hline
priordefects & probability of defects &  &  \\\hline
{[}0, 5) & 11.9\% &  &  \\
{[}5, 18) & 28.4\% &  &  \\
{[}18, 1588{]} & 59.6\% &  &  \\\hline
\multicolumn{4}{c}{\textbf{For release Chrome 41: 39.8\% of observations had defect}} \\\hline
priordefects & probability of defects & allchangescount & probability of defects \\\hline
{[}0, 4) & 16.0\% & {[}1, 10) & 13.5\% \\
{[}4, 18) & 33.5\% & {[}10, 48) & 33.8\% \\
{[}18, 1660{]} & 73.7\% & {[}48, 2415{]} & 73.4\% \\\hline
\multicolumn{4}{c}{\textbf{For release Chrome 42: 38.8\% of observations had defect}} \\\hline
priordefects & probability of defects &  &  \\\hline
{[}0, 4) & 15.3\% &  &  \\
{[}4, 17) & 34.9\% &  &  \\
{[}17, 1649{]} & 69.2\% &  &  \\\hline
\multicolumn{4}{c}{\textbf{For release Chrome 43: 28.8\% of observations had defect}} \\\hline
minorauthors & probability of defects &  &  \\\hline
0 & 19.1\% &  &  \\
{[}1, 108{]} & 71.4\% &  &  \\\hline
\multicolumn{4}{c}{\textbf{For release Chrome 44: 30\% of observations had defect}} \\\hline
allchangescount & probability of defects & minorauthors & probability of defects \\\hline
{[}1, 13) & 8.1\% & 0 & 19.6\% \\
{[}13, 52) & 23.7\% & {[}1, 96{]} & 75.9\% \\
{[}52, 1925{]} & 59.4\% &  &  \\\hline
\multicolumn{4}{c}{\textbf{For  Chrome - Combined: 35.6\% of observations had defect}} \\\hline
priordefects & probability of defects & minorauthors & probability of defects \\\hline
{[}0, 4) & 13.8\% & 0 & 25.2\% \\
{[}4, 18) & 32.0\% & {[}1, 108{]} & 76.1\% \\
{[}18, 1702{]} & 63.1\% &  &  \\\hline
\multicolumn{4}{c}{\textbf{For  All releases combined: 26.9\% of observations had defect}} \\\hline
priordefects & probability of defects & minorauthors & probability of defects \\\hline
{[}0, 2) & 7.4\% & 0 & 18.9\% \\
{[}2, 8) & 23.1\% & {[}1, 108{]} & 69.5\% \\
{[}8, 1702{]} & 54.6\% &  &  \\\hline
size & probability of defects &  &  \\\hline
{[}0, 113) & 7.3\% &  &  \\
{[}113, 571) & 21.9\% &  &  \\
{[}571, 106595{]} & 50.3\% &  & \\
\hline
\end{tabular}%
}
\end{table}

Table~\ref{t:MB_release} shows the variables that directly influence the \textit{defects} node in the Markov Blanket. Of the 11 review measures one is present in Qt 50, \emph{changesnodisc}, and another, \emph{allreviews}, is present in Chrome release 39. 
In Table~\ref{t:cpt}, we see the probability of a defect given each the review measure. A directory with two or more changes that are reviewed without discussion increases the probability of post-release defects by 42.3\% in Qt 50. For Chrome 39, the more often a directory is reviewed the more often there are defects and in the case of 36 or more reviews for a directory the probability of observing a defect increases to 69.8\%.
%n terms of how many times each variable appear in Markov Blanket of defects (for individual releases), we have:

%     allauthors allchangescount      allreviews   changesnodisc    minorauthors    priordefects    reviewissues            size
 %             1               2                            1               1                                   2               5                          1                     1
Of the 10 non-review ``traditional" measures six are present in one or more releases, see Table~\ref{t:MB_release}. \emph{allauthors,size} are present in one release each. \emph{allchangescount}, and \emph{minorauthors}, are each in $2/8$ releases. \emph{priordefects} the most common predictor is present in $5/8$ releases.
In Table~\ref{t:cpt}, we see the probability of a defect given each measure. 
We observe that all the predictors have a positive impact on
defects, \ie larger numbers increase the probability of
defects. Below we discuss ``traditional'' measures that are present in more than one release.
%minorauthors
Having one or more minorauthors that modify the files in a directory can also drastically increase the number of post-release defects by over 70\%.  
%allchangescount
Moderate changes to a component, \eg $\text{allchangescount}< 50$, can increase the number of defects by up to 33.8\%. Components with many changes, $ > 50$ can see post-release probabilities increase by over 60\%. 
%priordefects
The relationship between \emph{priordefects} is clearly shown in the CPT tables with a large number of prior defects, increasing the probability of post-release defects by between 30.6\% and 73.7\%.

\subsection{Indirect relationships and Visual Representation}

\begin{figure*}
%\centering
\includegraphics[width=\textwidth]{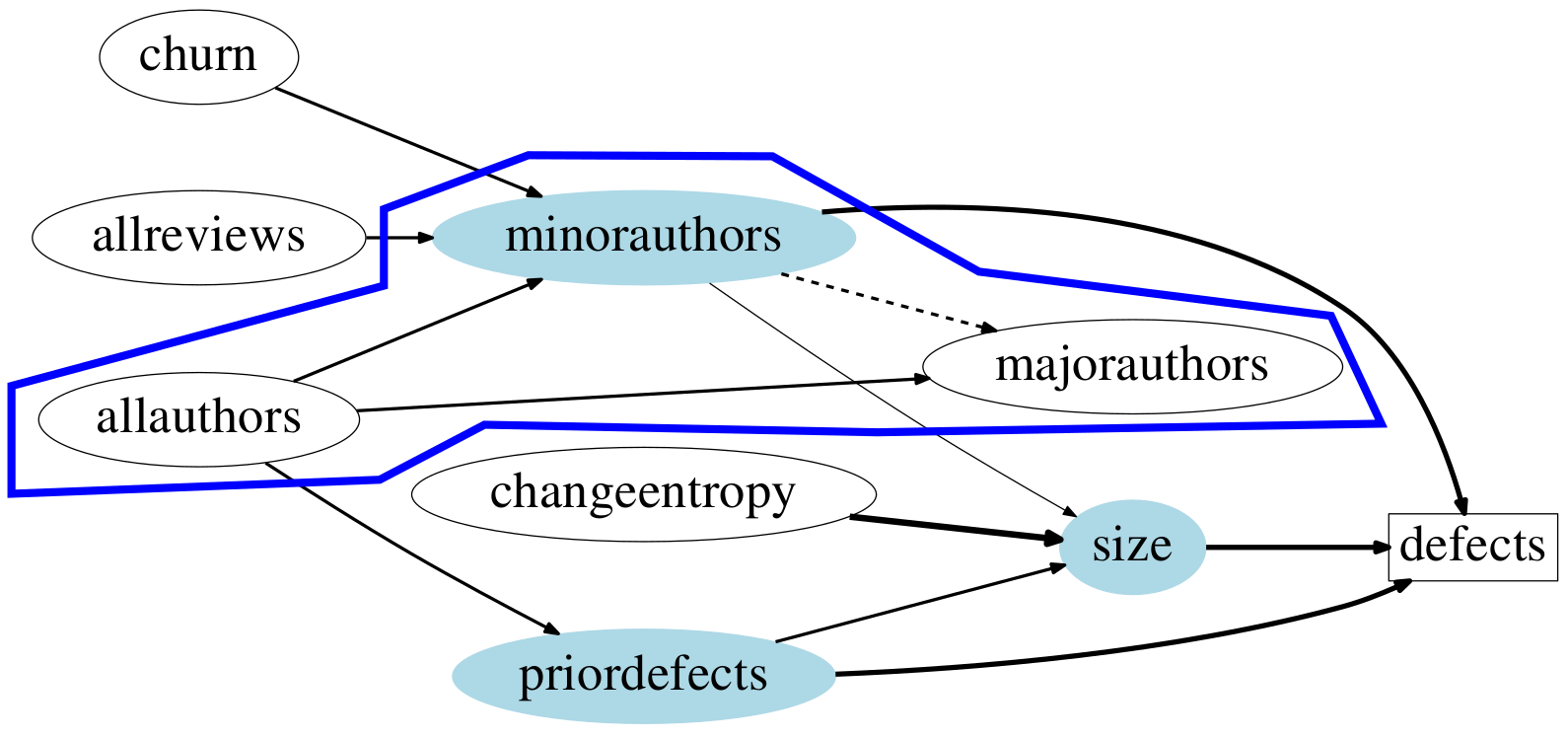}
\center
\caption{The Bayesian network graph for the combined Chrome and Qt data. The model confirms that review measures only indirectly impact post-release defects. Traditional measures, such as prior defects, have direct impact.}
\label{fig:model}
\vspace{-5pt}
\end{figure*}

In Figure~\ref{fig:model}, we present a snapshot of the BN model with only the \textit{defects} node, the nodes in its Markov Blanket, and the nodes that directly affect the nodes in the Markov Blanket for ease of interpretation, since the complete BN models with all nodes are rather complicated. However, the complete BN models for individual Chrome and Qt releases, the aggregate Chrome and Qt datasets, and for the combined dataset are available in our replication package~\cite{ReplicationPackage}.

%This graph allows us to see not only the measures that directly influnce post-release defects, but also the measure that have an indirect influence. 

The dotted edges indicate that the coefficient 
is negative for that edge, \ie increasing
the value of the parent node decreases the value of the child node
and vice versa. The immediate parents of the \textit{defects} node
(consequently, the Markov blanket for the \textit{defects} node in this case)
are colored in light blue and the \textit{defects} node is denoted in a rectangular shape.  The effect of each individual variable on the \textit{defects} node is shown in Table~\ref{t:cpt}. We do see a review measure: \textit{allreviews} affecting the variable \textit{minorauthors} in the BN, but it has no \textit{direct} impact on the number of defects in our combined model.

%\todo{add a more complete discussion -- added a few lines: Tapajit}

%\subsection{Comparing the result with the result from traditional modeling approach in Section~\ref{secReplication}.}
The variables indicated as the most important in the traditional modeling approach are also
the ones indicated as most important in BN modeling approach, except in the traditional model
the \emph{minor authors} variable was discarded as being redundant, and \emph{all authors} was used 
instead, while in BN approach \emph{minor authors} have a direct
influence on \emph{defects}. This illustrates the ability of BNs to
address some of the issues posed by correlated predictors, a situation common in software
engineering. 

The results from the two approaches are largely consistent in terms of indicating which variables 
are most significant in explaining the post release defects, and both approaches show that review-related
measures have no direct influence over the post release defects
variable in the combined model and no review measures are
statistically significant in more than one individual release. By
obtaining the same result
using two  
completely independent modeling approaches increases our confidence in the findings. Our conclusion to RQ 3, Structure of the relationships, is summarized in the following box.

\vspace{+5mm}
\fbox{ 
\begin{minipage}[b]{.9\textwidth}
{\bf Conclusion 3:} {\it Only prior defects, module size, and minor
  authors have direct effects on (form a Markov blanket for)
  post-release defects in the combined model }
 \end{minipage}
}
\\

\subsection{Addressing the issue of highly correlated variables --- problem of subjectivity in variable selection.} 
We have claimed before that BN modeling approach is not affected by the presence of highly correlated 
variables, and that can be seen in our BN model as well. The three author related variables: 
\emph{all authors}, \emph{minor authors}, and \emph{major authors} were highly correlated in our data. Therefore,
in the traditional modeling approach only one of them, \emph{all authors} was used in the final model.

In the combined BN model, the three variables appear connected to each other, as can be seen in 
Figure~\ref{fig:model} (the three nodes are inside the blue dotted polygon). The relationship
among the nodes from the BN model is easily interpretable: more \emph{allauthors} implies 
more \emph{minor authors} and \emph{major authors}, while increase in
\emph{minor authors} inevitably decreases 
\emph{major authors} as \emph{all authors} is the sum of minor and major 
authors. The BN model also suggests that the \emph{minor authors} variable has
substantially more influence over \emph{defects} than  \emph{major authors} or \emph{all authors},
thus it resolves the subjectivity in the variable selection problem.

To illustrate the usefulness of BNs it is worth making a few
additional observations. The size of the module tends to be
associated with code smells, effort, and defects (see,
e.g.,~\cite{SYAMD13,NMB08,AB06,M10}).  Not surprisingly, size affects
both prior defects and defects, since relative module size tends to
be stable release to release. 
 
More interestingly, the BN model in Figure~\ref{fig:model} suggests
that, for example, the presence of \emph{minor authors} both, increases the
size of the module (perhaps via unnecessary code bloat), and also
has a direct effect on the number of post-release defects (perhaps
due to lack of understanding of the module). Thus it has a double
effect on defects: direct, and mediated via module size. 

The variable \emph{minor authors} is, in turn, affected by the total number of
authors, the number of changes made to the module, and the number of
review issues. Arguably, the arrow should be pointing towards the
review issues from minor authors as it is the minor authors that are
likely to submit problematic code or be screened more vigorously
during the review (see more discussion on incorporation prior
knowledge in Section~\ref{sec:limitations}). However all these three
relationships (except the direction of the third, which can be
addressed by introducing a suitable prior) are rather intuitive.

Finally, it is worth considering the most important predictor of
defects: prior defects. Apart from size, it is also related to the
proportion of changes with no discussion, suggesting less aggressive
reviews, the typical number of reviewers, and, surprisingly, is
better for more complex modules. As noted earlier, the arrows should
arguably be reversed: modules with prior defects probably invite
more scrutiny with a larger review team. Why the modules with 
larger review teams tend to have higher proportion of changes with no
discussion may be worth a further investigation.    
\\

\vspace{+5mm}
\fbox{ 
\begin{minipage}[b]{.9\textwidth}
{\bf Conclusion 4:} {\it BN's can help address some difficulties
  posed by correlated predictors and the generative models help 
  articulate potential mechanisms of how development process and
  product measure interact.}
 \end{minipage}
}
\\

\section{Limitations} 
\label{sec:limitations}

In this section, we discuss factors that in our opinion may pose a threat to validity of the results we present. We inherit validity threats from the study we replicate and discuss new threats related to BNs. 

\subsection{External validity.}
McIntosh \etal~\cite{mcintosh2015emse} mined Android,
LibreOffice, QT, ITK, and VTK, but in the end the only project with enough bugs and links to reviews for a reasonable analysis was QT. Given the instability of the predictors and the difficulty in linking reviews on projects, we decided to use a new modelling framework on a single large successful project, Google Chrome, instead of a broad study of the predictors across multiple projects. While single case studies have value~\cite{Runeson2009EMSE,Menzies2013TSE}, clearly our results do not generalize beyond these two projects.

\subsection{Regression model}
We remove highly correlated variables and use the same model as McIntosh~\etal~\cite{mcintosh2014impact}. We do not consider interactions among variables because the model was already unstable and the additional complexity would further reduce stability in variable selection.

\subsection{Latent variables.}  Dealing
with hidden variables in Bayesian Networks remains an open research
question and an inherent limitation to all modeling techniques dealing with real observational data. However, this problem is not a serious threat to our results,
since we do not attempt to establish any causal relationship among the variables.
Our assumption to exclude potentially relevant unobserved
variables is ameliorated by the use of prominent predictors of software
defects used in extensive prior research on the subject.

\subsection{Discretization.} We transform our count variables to discrete variables using the Equal Frequency method as discussed in Section~\ref{secDiscretizationMethod}, and use two or three levels, based on the distribution of the original variables, for our discretized variables for the sake of simplicity in our final model. No discretization method is optimal, and the choice of the number of levels has subjectivity. However, as can be seen in the with the conditional probabilities in Table~\ref{t:cpt} the interpretations and bins seem reasonable given the software engineering context. 

\subsection{Threshold.} In order to obtain the final structure from
averaged model we use an arbitrary threshold of
confidence. We verify the robustness of the network
by gradually reducing the threshold and plotting the new
structure. The conclusion of the sensitivity analysis is that the
overall structure remains stable. In particular, the Markov
Blanket of \emph{defects} variable remains unchanged even for a
threshold value of 0.45. 
%Although our 
%analysis show that structure has quite low sensitivity to changes 
%in threshold we still list this as a threat because we do not have
% a good theoretical reason behind selecting that or another threshold value.  
 
\subsection{False Positive/Negative edges in the Bayesian Network.}
We used the best performing BN structure method as reported in~\cite{dey2020deriving}, 
and the final models were constructed based on repeating the search process 100 times.
However, there is still the possibility of false positive or negative edges in the model, 
but the impact of it on the final result is unlikely to be significant.

\subsection{Prior knowledge in BN structure search.}
We do not use any prior knowledge of the problem domain while learning the
BN structure. For instance, we have some prior knowledge about the
directions: \eg the \emph{defects} node should not have any outgoing
edges since it is measured after the release, and the \emph{prior defects}
node should not have any incoming edges since this information is
known a-priori.   This knowledge can be incorporated into the search process by providing the initial partial structure as a parameter for the search function.
Our unrestricted structure search yielded a model
where the first assumption does hold, but second one does not. There
is room for an argument that incorporating this prior knowledge
will result in a more realistic model, but a counter-argument may
be made as well. For example, in our model \emph{prior defects} might
represent a proxy measure for the inherent defectiveness of the
module, and using the assumed prior knowledge would have excluded
this possibility. Since this analysis was primarily concerned with
direct effects on defects and all the discovered links were pointing
inward (rendering the question moot), the ways to specify and
incorporate expert knowledge while being important by itself is
beyond the scope of this analysis.

\section{Discussion of Related Work}

In 1976 Fagan published the first empirical evaluation of software review, \ie inspection~\cite{Fagan1976IBM}. The work quantified the defect finding effectiveness of inspection based on the number of defects found per thousand lines of source code (KLOC) and percentage of total defects found by inspection. On the IBM system under study 38 defects per KLOC were found by inspection vs 8 per KLOC found by unit tests. Inspection found 82\% of the total defects found for the released product. In the intervening 40 years, code review has changed dramatically from the rigid inspection process that Fagan introduced.

Most of the early work on inspection focused on minor variations in the inspection process but kept the formality, measurability, and rigidity intact\cite{Martin1990ACM,Knight1993ACM,Kollanus2009OSEJ,Laitenberger2000SS,Wiegers2001AW}. The most important finding was that the inspection meeting need not be held in person to find a substantial number of defects~\cite{Votta1993SEN,Eick1992ICSE,Perry2002IEEE}. This lead the way to online review tools that ultimately lead to the currently popular and widely studied Gerrit~\cite{Mukadam2013MSR,mcintosh2015emse} and the pull request mechanism of GitHub~\cite{Yu2015WCRE,Rahman2014MSR,Gousios2015ICSE}.

There is also a long history of examining the factors that make peer review effective. 
Porter \etal
\cite{Porter1998ACM} examined both the process and the inputs to the process
(\eg reviewer expertise, and artifact complexity). In terms of the number of
defects found during review, Porter \etal concluded that the best predictor was the level of
expertise of the reviewers. Varying the processes had a negligible impact on
the number of defects found. This finding is echoed by others (\eg
\cite{Sauer2000IEEE,Kollanus2009OSEJ}).

Rigby \etal~\cite{Rigby2008ICSE,rigby2012contemporary,rigby2011understanding,Rigby2014TOSEM} examined open source software based review on multiple projects including the Linux kernel, the Apache server, and KDE. They created regression models with the number of defects found during review and the amount of time take for review. They found remarkably similar practices across project that had very little process, but relied on expert reviewers frequently reviewing each commit. In a study at Microsoft and AMD, Rigby and Bird~\cite{rigby2013convergent} found that these lightweight review practices were also used in industry. They also found that the focus had shifted from a defect finding activity to a problem solving one. 

Recent works have focused on the non-defect finding benefits of code review. For example, interviews of Microsoft and OSS developers have been conducted to understand developer motivations for code review~\cite{bacchelli2013expectations,Bosu2017TSE}. They found that while developers want to find defects, they were also interested in spreading knowledge and discussing alternative solutions. Indeed, code review has also been shown to be effective at spreading knowledge and reducing the impact of code ownership~\cite{rigby2013convergent,kononenko2015investigating,Thongtanunam2016ICSE}. Other works focused on the types and utility of feedback provided by developers~\cite{Bosu2015MSR,Beller2014MSR,Kononenko2016ICSE} and on the ability of code reviews to identify security vulnerabilities~\cite{Bosu2014FSE,Meneely2017EMSE}

Despite these additional benefits of code review, the primary goal is still defect finding~\cite{bacchelli2013expectations,Bosu2017TSE}. The literature abounds with papers that use product and process metrics to predict where defects will occur, for example,~\cite{Fenton1999TSE,Neuhaus2007CCS,Shivaji2013TSE}. These models have also been used to understand changes in development practices, such as co-location vs remote developers~\cite{bird2011don}, the impact of developer turnover~\cite{M10} and much more. As far as we know, McIntosh \etal's~\cite{mcintosh2015emse,mcintosh2014impact} is the first to examine to include peer review measures into a defect model. Earlier works ~\cite{Porter1998ACM,Rigby2014TOSEM} measured how many defects where found during the review, but did not look at the long-term impact of review on defects. As a result, our work first replicates McIntosh \etal's work that covered only releases (two Qt releases, one release from ITK, and one from VTK), we expand the study to include six releases of the Chrome project.

 A case for use of BNs in the context of Software Engineering was made
by Fenton et.al.~\cite{fenton1999critique,fenton2002software}, while the earliest publications
utilizing BNs we could find~\cite{HM03a} constructed search of the
structure based on the statistical significance of partial
correlations in the context of modeling delays in globally
distributed development. \cite{stamelos2003use,pendharkar2005probabilistic} considered
the application of Bayesian networks to prediction of effort, 
\cite{fenton2007predicting,neil1996predicting,okutan2014software} 
used Bayesian networks to predict defects, and \cite{pai2007empirical} 
used BN approach for an empirical analysis of faultiness of a software. 
In a similar work, ~\cite{bai2005bayesian} used modified BNs (Markov Bayesian network ) 
for software reliability prediction.~\cite{van2006application} used BNs for predicting
 maintainability of Object Oriented software, and~\cite{bibi2003bayesian} used BNs 
as a software productivity estimation tool. We are not aware of
prior applications of Bayesian Networks for modeling software reviews.
On the other hand, Bayesian structure learning is a big domain in itself 
with a wide range of algorithms, but its use in software engineering context 
is not very common.

\subsection{Conclusion} 
\label{section_7}

Prior works have shown that the  defects are both effectively and efficiently
found during code review \cite{fagan2002history,Porter1998TOSEM,Rigby2014TOSEM}. Recent works provided qualitative evidence that reviews
provide benefits beyond defect detection, such as knowledge sharing
\cite{Sauer2000TSE,bacchelli2013expectations,rigby2011understanding,Bosu2015MSR,Kononenko2016ICSE,Rahman2017MSR}. In contrast, the goal of this
work is to understand if code review measures can quantify the longterm impact of peer review on post-release defects.

 \subsubsection*{\large Conclusion 1: Reproduction and Replication}

\leavevmode \\
% we replicate McIntosh
McIntosh \etal \cite{mcintosh2014impact,mcintosh2015emse} were the first to study the impact of code review measures on post-release defects. We replicated their study using data they provided and as well as on the Chrome data we extracted. We discussed our findings with the first author of the original study.
%different results
McIntosh \etal found that review participation had an influence on post-release defects, but we were unable to replicate these results. 
%find that ...
Instead we found that review measures contributed little to the performance of the model. The $R^2$ values with and without review measures were almost identical. 
In agreement with existing defect prediction work \cite{mockus2000case,hassan2009predicting,bird2011don,graves2000predicting}, our results show that prior defects, the module size, and the number of authors are the strongest predictors of post-release defects. Review measures are neither necessary nor sufficient to create a good defect prediction model.

\subsubsection*{\large Conclusion 2: Inconsistent Models} 

\leavevmode \\
It is extremely difficult to replicate an empirical software study that
involves both mining operational data and statistical modelling.  Despite using
exactly the same data and modelling approach we obtain substantially different
results.  
In both our study and that of McIntosh \etal~\cite{mcintosh2015emse} a key
problem is the need to select an uncorrelated set of variables. The variable
selection process is inherently subjective because differences in expert opinions
may lead to different sets of variables. 

Furthermore, in both studies, the models were performed per project and per release. Even strong predictors, such as prior-defects varied substantially in their predictive power between project releases. This result suggests an issue with the traditional variable selection used in regression models.

\subsubsection*{\large Conclusion 3: Direct effects}
\leavevmode \\
Regression models require the researcher to define a response and a set of predictors. This approach lacks tools to distinguish between an actual relationship and the effect of a shared confound. In contrast, Bayesian Networks remove the need for variable selection and shows the Bayesian relationships among variables.
The term ``direct effect'' is  meant to  quantify an influence that
is  not mediated by other variables in the  model  or, more
accurately, the sensitivity of $Y$ to changes in $X$ while all other
factors in the analysis are held fixed. Indirect effects can
manifest themselves on the response only through affecting the value
of predictors that gave direct effects on the response. 

According to our BN, only three measures directly impact post-release defects: the number of prior defects, the number of minor authors, and the size of the module. The code review measures, such as \emph{rushed reviews}, \emph{number of review participants}, and \emph{discussion length}, did not directly impact the number of post-release defects.

\subsubsection*{\large Conclusion 4: Generative models and indirect effects}

\leavevmode \\
The use of BN provides a way to evaluate the indirect effects
that code reviews have on defects through the influence on other
variables. Such indirect effects bedevil traditional analysis
methods that use observational data. If the set of observed
variables is complete, it is possible to calculate an impact of
intervention akin to the results that could be obtained only in
randomized experiments. For example, changes that have \emph{no review
discussion} tend to be associated with files that have had many
\emph{prior defects} which in turn increase the number of post-release
defects.  
A further example from our BN model shows that having 5 or more
reviewers is seen to increase chance of having post-release defects
from 20\% to 33\% through mediating variables \emph{allauthors} and \emph{minorauthors}.

We have demonstrated the difficulties in using traditional models on
observational data. Although individual code reviews find defects, we were
unable to find any direct effect of review measures on post-release defects. By
using BN we found that code review measures indirectly effect post-release
defects. We hope that other researchers will use the approaches presented here
to untangle the relationships among software measures. These
indirect effects should provide a more nuanced understanding of
software engineering. We make our scripts and data available in our replication package~\cite{ReplicationPackage}.

%\begin{acknowledgements}
%\end{acknowledgements}

\bibliographystyle{abbrv}
%\bibliographystyle{ACM-Reference-Format}
\bibliography{vm_ref} 
%\begin{thebibliography}{}

% etc
%\end{thebibliography}

\end{document}